\documentclass[preprint2]{aastex}
\shorttitle{ 
'Radio Sources in RCS1 Galaxy Clusters'  } 
\shortauthors{ 
Gralla & Gladders}
\usepackage{graphicx}
\usepackage{amssymb}

\usepackage{epstopdf}
\usepackage{epsfig}

\begin{document}

\title{Constraining the Redshift Evolution of FIRST Radio Sources in RCS1 Galaxy Clusters}
\author{
Megan B. Gralla\altaffilmark{1,2},
Michael D. Gladders\altaffilmark{1,2},
H. K. C. Yee\altaffilmark{3} and
L. Felipe Barrientos\altaffilmark{4}
}

\altaffiltext{1}{Department of Astronomy \& Astrophysics, University of Chicago, 5640 S. Ellis Avenue, Chicago, IL 60637}
\altaffiltext{2}{Kavli Institute for Cosmological Physics, Department of Astronomy \& Astrophysics, University of Chicago, 5640 S. Ellis Avenue, Chicago, IL 60637}
\altaffiltext{3}{Department of Astronomy and Astrophysics, University of Toronto, 50 St. George St., Toronto, Otario, M5S 3H4, Canada}
\altaffiltext{4}{Departamento de Astronom\'ia y Astrof\'isica, Pontificia Universidad Cat\'olica de Chile, Vicu\~na Mackenna 4860, 7820436 Macul, Santiago, Chile}

\begin{abstract}
We conduct a statistical analysis of the radio source population in galaxy clusters as a function of 
redshift by matching radio sources from the Faint Images of the Radio Sky at Twenty-Centimeters (FIRST) 
 catalog with 618 optically-selected galaxy 
clusters from the first Red-Sequence Cluster Survey (RCS1).  
The number of excess radio sources (above the background level) per cluster is $0.14 \pm 0.02$ for clusters
with $0.35 < z < 0.65$ and is $0.10 \pm 0.02$ for clusters with $0.65 < z < 0.95$.
The richest clusters in the sample have more radio sources than 
clusters with low or intermediate richness.  When we divide 
our sample into bins according to cluster richness, we do not observe any significant difference ($> 1.5\sigma$) 
in the number of radio sources per unit of cluster mass for the galaxy clusters with $0.35 < z < 0.65$ as compared to the 
galaxy clusters with $0.65 < z < 0.95$.  
Thus the entire sample can be characterized by the number of (L$_{1.4 GHz} > 4.1 \times 10^{24}$ W Hz$^{-1}$) radio sources per unit (10$^{14}$ M$_{\odot}$) mass, which we measure to be 0.031$\pm$0.004.
We further characterize the population of galaxy cluster-related radio sources through visual 
inspection of the RCS1 images, finding that although the radio activity of brightest cluster galaxies (BCGs) also does 
not strongly evolve 
between our high and low redshift samples, the lower-redshift, richest clusters are more likely to host
radio-loud BCGs than the higher-redshift, richest clusters or poorer clusters at the 2-$\sigma$ level.  
\end{abstract}

\keywords{galaxies: clusters: general --- galaxies: active}

\section{Introduction}

Historically, observing the environments of powerful radio sources has been a useful method of finding 
high-redshift galaxy clusters \citep[e.g.,][]{radioclustersearch}.  Also, much detailed study has been done
of the radio source population in nearby galaxy clusters \citep[e.g.,][]{ledlow1,owenled,ledlow}. 
Now that uniformly selected samples
of more distant galaxy clusters exist, we can address the redshift evolution of the population of 
radio sources in galaxy clusters; that is, given a complete
sample of galaxy clusters, how many of them contain powerful radio sources as a function of cosmic time? 
The answer has implications for galaxy formation, 
 galaxy cluster physics and the selection function of 
large SZ-selected galaxy cluster surveys.

Heating from powerful radio jets of AGN is one feedback mechanism invoked to explain the observed 
X-ray properties of the centers of galaxy clusters \citep[][for a review]{binneytabor, ccreview}.  
Studies such as \citet{burns}, \citet{eilek} and \citet{mittal}  have provided evidence that most (and maybe all) galaxy clusters with
X-ray cooling cores also contain powerful radio-loud AGN in their brightest cluster galaxies (BCGs),
suggesting that the radio activity of the black hole of the central galaxy is linked to the thermodynamics
of the cluster gas near the core.  
Thus, understanding how the population of radio-loud AGN is related to cluster mass and how this 
population evolves  with redshift is important to understanding the energy budgets in the gaseous 
cluster cores.  Radio-loud AGN feedback may also be important in galaxy evolution.  For example, 
applying semi-analytic models to the output from the Millenium Simulation, \citet{croton} found 
that not only do radio AGN solve the problem of the over-cooling
of gas in the cores of galaxy clusters, but that the feedback from radio AGN affects the colors, morphologies and 
stellar ages of the massive galaxies in clusters, bringing them into better agreement with observations
than models without such feedback.  

Current experiments such as the South Pole Telescope and the Atacama
Cosmology Telescope 
use the Sunyaev Zel'dovich (SZ) effect to discover galaxy clusters
\citep{sptmoreclusters, actclusters}.  These experiments operate at millimeter and sub-millimeter wavelengths and 
take advantage of the SZ effect's nominal independence from cluster redshift.  While most 1.4 GHz radio 
sources have little power at such high frequencies, a population of radio sources with flat or 
inverted spectra could contaminate these surveys \citep[see ][ for an investigation of the spectral properties of cluster radio sources]{bruce}.  
Little is currently known about the redshift 
evolution of the radio sources associated with galaxy clusters, but it could affect the cluster 
sample selection as a function of redshift for these surveys \citep{linmohr}; investigating the 
redshift evolution of the radio source population in clusters even at 1.4 GHz provides a useful 
constraint to high frequency experiments.

In this paper, we study the population of radio sources from the Faint Images of the Radio Sky at Twenty-Centimeters
(FIRST) survey that lie 
within the projected inner 0.5 Mpc of galaxy clusters from the first Red-Sequence Cluster Survey \citep[RCS1;][]{rcs1}.  The 
optically selected RCS1 galaxy cluster sample is well-characterized over a wide range in redshift \citep[0.35 to 0.95;][]{rcscosmology}, 
allowing us to systematically address the evolution of the galaxy clusters' 
radio source population using a uniformly selected sample.  

Throughout this work we assume a flat  $\Lambda$CDM cosmology with WMAP-5 parameters: $\Omega_{m} = 
0.273$, $\Omega_{\Lambda} = 0.726$, and $h = 0.705$ \citep{wmap5}.  

\section{Data}

\subsection{Red-Sequence Cluster Survey}

The RCS1 data cover a total of about 90 square degrees in the R$_{C}$ and $z'$ filters.  The data were
taken at the Canada-France-Hawaii Telescope (CFHT) and the Cerro Tololo Inter-American Observatory 
(CTIO) 4m.  \cite{rcs1} describe the details of the data acquisition, data reduction and galaxy 
cluster survey generation and characteristics.  The survey is designed to take advantage of the red 
sequence cluster detection algorithm \citep{rsalgorithm}.  

In this sample, galaxy cluster centers are determined by the center of light from red galaxies, and 
redshifts for the galaxy clusters are computed using observed red-sequence colors.  
The redshifts are typically accurate to better than 0.05 for $0.2<z<1.0$, as indicated by follow-up spectroscopy \citep{rcsspec}
and modeling of the RCS data \citep{rcscosmology}.
See \cite{rcs1} for 
details on the derivation and uncertainties on these properties.  The cluster richness parameter B$_{gc}$ 
is used as a mass proxy and is defined as the amplitude of the galaxy$-$cluster center correlation 
function, which scales as the number of galaxies within 0.5 Mpc (evaluated for $h=0.5$) of the cluster 
center normalized by the galaxy luminosity function and spatial distribution.
\cite{bgcpaper} show that B$_{gc}$ scales well with other mass proxies using the CNOC 
cluster sample.   They empirically derived the following relation based on dynamically determined mass measurements 
in order to convert from richness to 
cluster mass, which we have scaled to give us $M_{200}$ for $h = 0.705$:
\begin{equation}
\log M_{200, h=0.705} = (1.64 \pm 0.28) \log B_{gc} + (9.90 \pm 0.89)
\end{equation}

We use B$_{gc,R}$ (calculated with $h = 0.5$), which is derived using galaxies in the red sequence.  B$_{gc,R}$ is expected to 
 trace the cluster mass over a wider redshift range than B$_{gc}$, as the red galaxies evolve slowly and the cluster red 
sequence was well in place by $z = 1$.  B$_{gc,R}$ is less likely to be affected by variations and evolution in the blue fraction of
cluster galaxies.
Indeed the cosmological analysis of the RCS1 survey does not detect evolution in the mass-richness relation \citep[although
the constraints are weak; ][]{rcscosmology}.
Because the above mass-richness relation was derived for B$_{gc}$ rather than B$_{gc,R}$, we have applied 
a correction factor of 1.23 to the B$_{gc,R}$ measurements when we calculate mass from richness.  We empirically
determined this factor by using both B$_{gc,R}$ and B$_{gc}$ for the 212 clusters 
in our sample which overlap with the redshift range of the CNOC clusters ($0.35 < z < 0.54$) and calculating the mean
ratio of the two richnesses.  See \citet{yeelopez} for a comparison between Abell richness class and B$_{gc}$ (the median
B$_{gc,R}$ of a cluster with Abell richness class 0 is about 1090 for $h=0.705$).

We impose a richness cut so that only galaxy clusters with B$_{gc,R} > 300$ are included.
  We also require that the uncertainty on the richness be less than half the richness value and that the 
significance of each cluster detection be greater than 3.3 $\sigma$.  This significance limit was applied in the cosmological study
of the RCS \citep{rcscosmology}, which found that for clusters above this limit, the contamination rate is $< 0.5\%$.
The redshift range for the clusters is restricted to 
0.35 to 0.95.  The resulting subsample of galaxy clusters is well-defined 
within a cosmological context over this broad redshift interval, as we are using the same sample constraints
used in the cosmological analysis of the RCS1 survey \citep{rcscosmology}.   

The combined area of overlap between the RCS1 and the FIRST surveys is about 40 square degrees.    
We limit the cluster sample so that it contains only galaxy clusters
that are within and further than $38 \arcmin$ from the edge of the FIRST survey, which is just greater than
10 Mpc at a redshift of 0.3.  Because we also limit our cluster sample
to clusters with $z > 0.35$, the 10 Mpc area surrounding each cluster in our sample
fully lies within the footprint of the FIRST survey (thus, we may use a 10 Mpc annulus around the clusters'
centers when calculating the background level of radio sources, see section \ref{method}).  
 
Constraining the RCS1 cluster catalog in this way yields a sample of 618 galaxy clusters. 

\subsection{FIRST Radio Survey}
The FIRST survey was conducted at the National Radio Astronomy Observatory (NRAO) Very Large Array (VLA) 
by \citet{becker}.  The survey consists 
of observations at 1.4 GHz.  The VLA was in its B configuration, giving the resulting maps a 
resolution of $~$5\arcsec.  The survey reached a point source sensitivity depth of $\sim$1.0 mJy.  Radio sources were 
selected and cataloged by \citet{white}.  Source positions are accurate to better than 1\arcsec, 
enabling matches to individual optical objects in many cases.  Because of the high resolution of the
survey data, radio sources that are extended, particularly radio sources with double lobe morphologies,
are often resolved and counted as multiple radio sources.  We address these multiple component radio 
sources in section \ref{multipleradiosources}.

\section{Analysis and Results}

\subsection{Statistical Analysis of Radio Sources around Clusters}

\subsubsection{Methods \label{method}}
Radio luminosity is calculated from flux using the following 
relation: $L_{1.4 GHz} = 4 \pi F_{1.4 GHz} {D_A}^{2} (1 + z )^{3 + \alpha}$, where we assume a spectral 
index of  $\alpha = 0.8$ where $L_{\nu} \propto \nu^{-\alpha}$. 
According to this relation, a radio source at the flux limit of the FIRST survey (1 mJy) and at the redshift
limit we are imposing on the RCS1 survey ($z = 0.95$) would have a luminosity of $4.1 \times 10^{24}$ W Hz$^{-1}$, so
we adopt this as our luminosity cutoff. 
 At this luminosity it is likely that radio emission is dominated by AGN activity rather 
than star formation, as radio sources dominated by star formation mostly lie below $10^{23}$ W 
Hz$^{-1}$ \citep{fieldrlf}.  
While there is some evidence that this crossover limit in the radio luminosity function evolves
with redshift, our sample is still well above the limit even at $z=1$ \citep{owentalk}.  
Also, it has been shown 
in nearby clusters that the crossover between the populations is fainter in clusters than in 
the field \citep{millerowen}, further ensuring that cluster radio sources brighter than the luminosity
limit we apply are dominated by AGN emission.

For each galaxy cluster in our sample, we identify all nearby FIRST
radio sources.   The cluster redshift is used
to determine the corresponding position (in proper megaparsecs) with respect to the cluster center and luminosity 
for all of these nearby radio sources.   
The luminosity cutoff determined above is then imposed so that only radio sources above this luminosity
limit are counted in this statistical analysis.  We repeat this procedure for every galaxy cluster in the
sample, compiling the list of positions and luminosities for all radio sources around all clusters.
Adopting the cluster redshifts for background and foreground radio sources does not bias our results in any way
because we use the resulting positions and luminosities of radio sources with respect to the clusters
to statistically determine the background level for the sample.  

Because of the high resolution of the FIRST data, extended sources (for example, two lobes of an AGN)
are split into multiple sources in the catalog.  These sources are thus counted twice in this 
statistical analysis.  Also, if a radio source is nearby two galaxy clusters, it will be counted twice,
once for each cluster.  We addressed these issues by visually inspecting the galaxy clusters that 
contain radio sources within a projected 0.5 Mpc; see section \ref{visual}.  While there were 16 galaxy clusters
with centers within 0.5 Mpc of another galaxy clusters as determined at the redshift of one cluster (7 of which
were pairs, 2 of which were within 0.5 Mpc of other clusters according to their redshifts but not according to the other
clusters redshifts), no radio sources overlapped between two different clusters.

\begin{figure}
\plotone{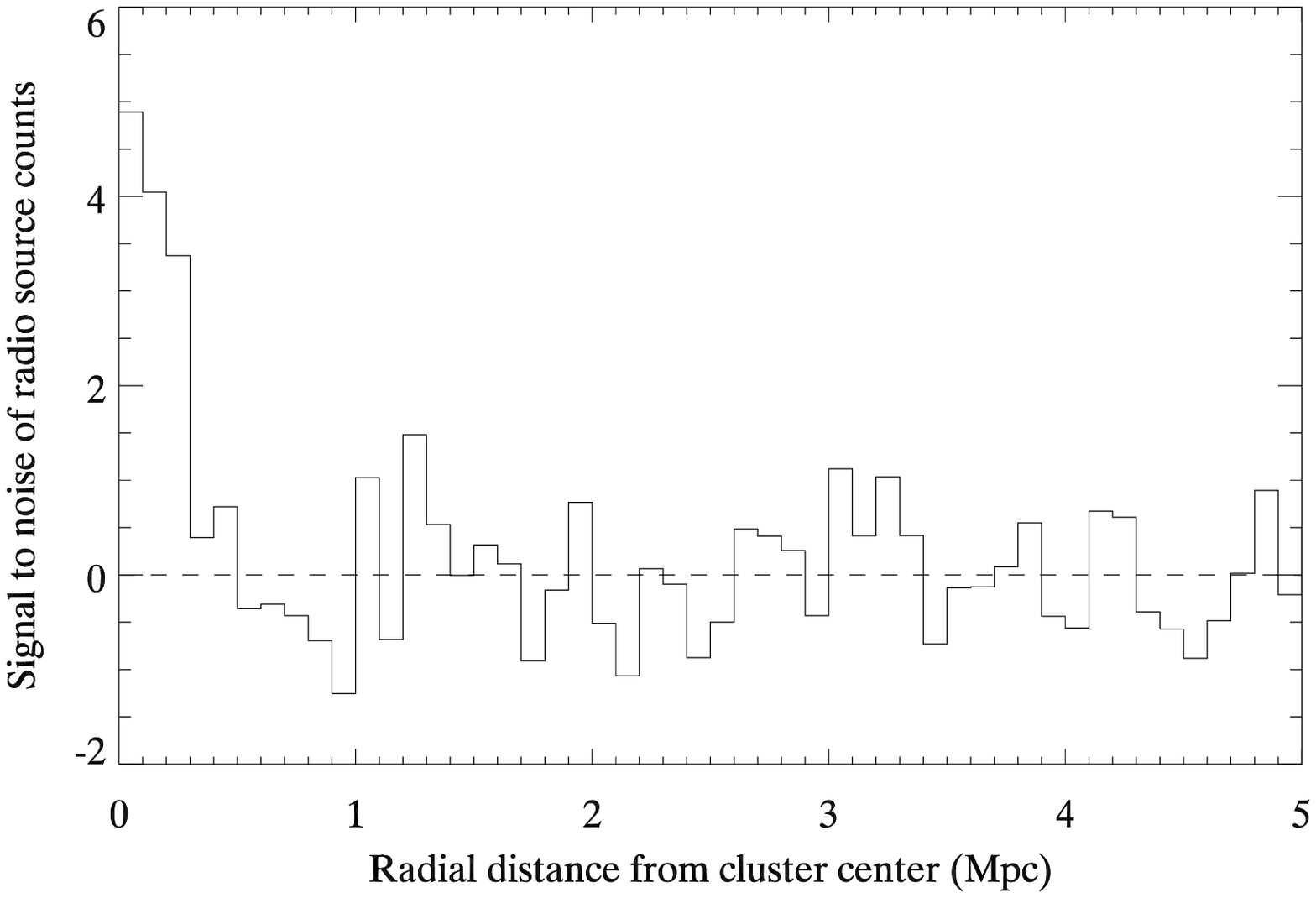}
\caption{The signal-to-noise ratio of each radial bin of the number of FIRST radio sources around RCS1 galaxy 
clusters, after a constant surface density background level has been subtracted.  The dashed line shows
the zero level at which the counts are equal to the background level.  The radio source positions are
calculated adopting the clusters' redshifts for the FIRST sources.
The background level is calculated in an annulus 5 to 10 Mpc from the cluster center. \label{radial}}
\end{figure}

In order to calculate the background level of radio sources, we determine the number of radio 
sources above the luminosity cutoff (with the radio source luminosities determined for the cluster redshifts
as discussed above) in a 5 to 10 Mpc annulus around the cluster centers.  We 
extrapolate this constant background
 surface density of radio sources to the inner cluster regions, subtracting the background 
contribution to calculate the number of excess radio sources.  
The uncertainties are calculated assuming Poisson statistics for both the background
and the signal.  
Figure \ref{radial} shows the signal-to-noise ratio in radial bins around the clusters' centers 
of the number of excess radio sources above the background level.
\placefigure{radial}

As can be seen in the figure, most of the excess radio sources are within 0.5 
Mpc of the cluster center, so we define our statistical sample of radio sources in galaxy clusters as the 
excess number of radio sources within a projected radius of 0.5 Mpc of the clusters' centers. 
  The total number of radio sources (including background sources) that we find within 0.5 Mpc 
of clusters' centers is 111, which is significantly higher than the background, from which we would 
expect $36.2 \pm 0.3$.  The cluster with the highest number of radio sources within 0.5 Mpc has 4, 
although 3 of these are multiple components of the same radio galaxy.  For a discussion of the frequency
of multiple radio sources (both independent and multiple components) in clusters, see section \ref{multipleradiosources}.

Correctly handling the background population of radio sources is important in this analysis because 
the lower redshift objects subtend more area of sky than high redshift objects, so poor background 
estimation could mock redshift evolution.  
We check our background estimation against a bootstrap analysis in which we calculated the number of 
radio sources around random positions within the area of overlap between RCS1 and FIRST.  
Restricting this analysis to the survey region ensures that we did not sample more or less sensitive 
regions of the FIRST survey than our data cover so that unevenness in radio coverage does not affect
 our background estimation.  We assigned each random position a ``redshift'' drawn from the actual
cluster redshifts of the galaxy clusters.
The background-subtracted number of radio sources per random position is consistent with zero.  This
remains true for both the random positions that were assigned redshifts less than 0.65 and for random
positions that were assigned redshifts greater than 0.65.
Thus, we see no evidence for large scale structures correlated on scales of 5 to 10 Mpc from the
clusters' centers biasing the background level we calculate, and we do not see any bias from 
the background calculation that could introduce discrepancies between the high and low redshift samples.

\subsubsection{Redshift Evolution\label{stats}}

The total number of radio sources within 0.5 Mpc of clusters' centers in our low redshift (0.35 $<$ z $< $ 
0.65) sample is 62 radio sources in 331 galaxy clusters, and in our high redshift (0.65 $<$ z $<$ 0.95)
 sample is 49 radio sources in 287 clusters.  The number of excess radio source components per cluster (after 
background subtraction) is 0.14 $\pm$ 0.02 for the low redshift sample and 0.10 $\pm$ 0.02 for 
the high redshift sample.  Errors are estimated assuming Poisson statistics in the signal and 
background number of radio sources.   
The significance of the difference between the high and low redshift samples is $1.4 \sigma$.
Thus, the level of evolution in this sample is constrained to be small (especially in comparison with the evolution 
in cluster X-ray AGN, see discussion in section \ref{xray}).

\begin{figure}
\plotone{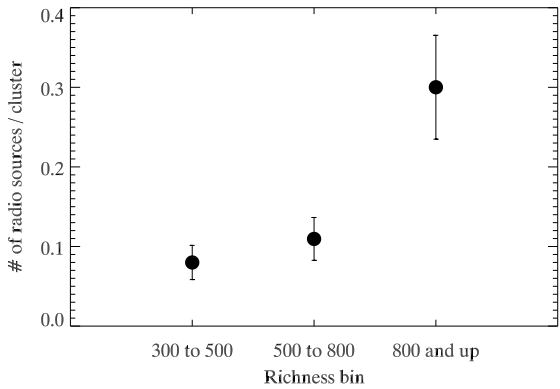}
\caption{The number of radio sources per cluster in three different richness bins 
(using the B$_{gc,R}$ richness parameter). 
   \label{bgcall}}
\end{figure}

\begin{figure}
\plotone{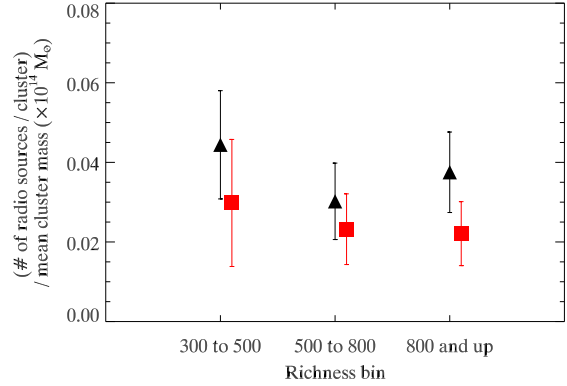} 
\caption{The number of radio sources per cluster, normalized by average cluster mass, derived from cluster richness,
 for each richness 
bin.  The average cluster masses are 2.1 $\times10^{14}$ M$_{\odot}$, 4.1 $\times10^{14}$ M$_{\odot}$, and 1.0 $\times10^{15}$ M$_{\odot}$ 
for the three richness bins. The black triangles correspond to the low redshift sample and the red squares correspond to the high 
redshift sample. \label{radz}}
\end{figure}

\begin{figure}
\plotone{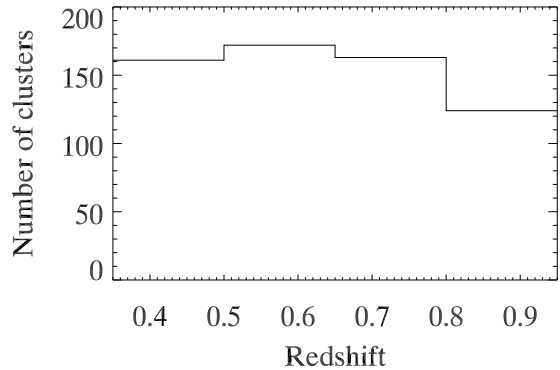}
\caption{Histogram of redshifts of the galaxy clusters in this study. \label{zhist}}
\end{figure}

\begin{figure}
\plotone{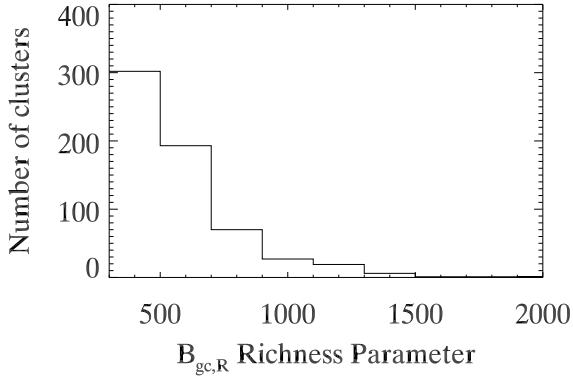}
\caption{Histogram of B$_{gc,R}$ richness parameters of the galaxy clusters in this study. \label{bgchist}}
\end{figure}

An important correlation we notice in our data \citep[which has been previously noted, see for example][]{bestbcgrich}
is that the richest clusters are more likely to host 
a radio source than poorer clusters (see Figure \ref{bgcall}). 
In order to investigate whether there are differences in the high
and low redshift radio source populations in clusters of similar mass, we split our galaxy clusters
into three bins according to cluster richness (Figure \ref{radz}).  
 The B$_{gc,R}$ values of 300, 500, and 800
correspond to masses of $1.3\times10^{14}$ M$_{\odot}$, $3.0 \times10^{14}$ M$_{\odot}$, and $6.4\times10^{14}$ M$_{\odot}$,
 respectively.  
There are 181 clusters with B$_{gc,R}$ between 300 and 500
at low redshift (0.35 to 0.65) and 121 at high redshift (0.65 to 0.95), there are 110 clusters with richnesses between 500 and 800 at low redshift
and 122 at high redshift, and there are 40 clusters with B$_{gc,R}$ 800 and up at low redshift and 44 at high redshift.  See Figures \ref{zhist} and
\ref{bgchist} for the distributions of the redshifts and richnesses of the galaxy clusters.
We note that the scatter in the mass-richness relation is considerable, so for any one cluster the mass is not well measured.

When we normalize the number of radio sources 
per cluster by the average mass, as derived from richness, for each richness bin,
the number of radio sources per unit of cluster mass is roughly constant, with a value of $0.031 \pm 0.004$ per 10$^{14}$ M$_{\odot}$.
Because the masses are derived from optical richnesses, this implies that the number of radio sources
per cluster galaxy is constant, as has been noted in previous studies \citep{ledlowrichnesses}.  Comparing the normalized 
excess counts of radio sources per cluster, we find that
the low-redshift clusters have more radio sources per unit mass than the high-redshift clusters at the 1.5$\sigma$ level.
This may be driven by a similar trend in the frequency of radio activity in the clusters' BCGs (see below, 
section \ref{bcgs}); however, the significance of the difference is low.  Alternatively, the data constrain the redshift evolution 
to be small.  The 1-$\sigma$ uncertainty on the number of radio sources per unit mass
for each bin in redshift and richness ranges
 from 27 to 53$\%$ of the signal, making evolution by factors of a few unlikely.

\subsection{Visual Inspection\label{visual}}

\subsubsection{Methods}
To supplement our statistical analysis, we visually inspected images from the RCS1 survey data of the galaxy clusters in our 
sample.  We did this for a number of reasons: to better investigate the relationship between the brightest cluster galaxies 
(BCGs) and the radio sources, to determine the frequency of multiple radio sources within 
single galaxy clusters, and to identify radio sources that are 
obviously associated with foreground or background galaxies.

When visually inspecting the cluster images, we considered the cluster redshifts and richnesses, and we
identified bright potential cluster members.  
Each cluster's center is defined as the center of red light, not a BCG position.
BCG candidates were identified according to the following
criteria: central position in the cluster, morphological appearance, position in the color-magnitude
diagram, and the presence of satellites or an extended halo of optical light.  
While in principle we could simply select the brightest galaxy within some
distance from the cluster center and with some color 
near the red sequence, we are more intereseted in the
central galaxy of the galaxy cluster, and in practice the identification of this can be ambiguous.  For example, there may be
two bright central galaxies where one is slightly brighter, but another is more central and has nearby
satellite galaxies.  Or the brightest, dominant galaxy could be offset from the cluster center, particularly for clusters with 
extended or disturbed morphologies.  Sometimes multiple candidates were identified as possibilities.

Radio sources were also visually inspected.  Radio sources were matched to optical
counterparts when possible, and radio morphology was noted.

\begin{figure}
 \plotone{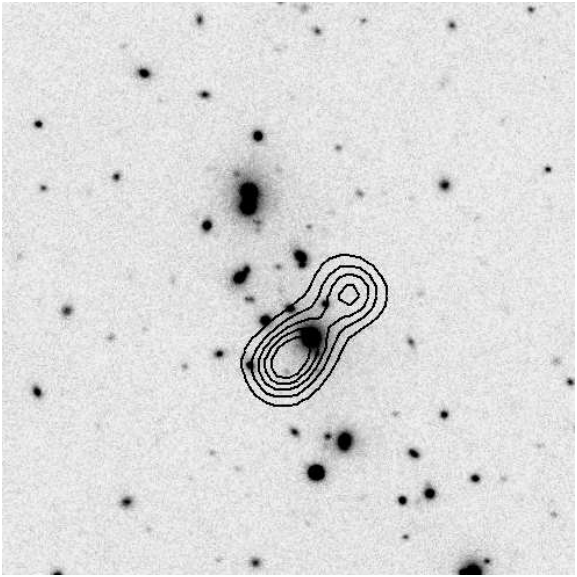}
\caption{RCS1 $z'$ band image of the galaxy cluster RCS1 J132655+302112, with radio contours from the 
FIRST image archive.  
The image is $0.5 \times 0.5$ Mpc across at
the cluster redshift of 0.37.  
The radio contours are at the 1 to 5 mJy levels, in steps of 1 mJy.
This cluster is an example of a multiple-component extended radio source that 
looks like it is associated with the cluster BCG.
\label{prettypic}}
\end{figure}

As an example of the benefits of this visual inspection of the cluster images, Figure \ref{prettypic}
shows the radio emission in the galaxy cluster RCS1 J132655+302112.  This
galaxy cluster is at redshift 0.37 and has a B$_{gc,R}$ of 1425 (corresponding to M$_{200} = 1.7 \times 10^{15}$ M$_{\odot}$).
In this example, two components of an extended radio source are identified as sources in the FIRST catalog.  These radio sources
 appear to be two lobes radiating from a bright central galaxy, a good BCG candidate located about
0.1 Mpc from the cluster center (as identified by the center of the red galaxies' light.)  Visual
inspection identifies this system, but simply matching individual radio sources with individual
optical objects using a small radius of tolerance could miss sources like these.  This example
demonstrates the usefulness of characterizing the population by eye in order to better understand the
 nature of the population.  Of course, both of the components of this radio source are within the
central 0.5 Mpc of the galaxy cluster, so this was easily identified in the statistical sample.

\subsubsection{Radio-loud BCGs\label{bcgs}}
Previous studies have shown that BCGs are more likely to be radio-loud than
 other galaxies of similar stellar mass \citep{bestbcgrich}, indicating a difference in the radio-loud
BCG population relative to the general cluster radio source population.  We investigate whether there is a 
redshift evolution or cluster mass dependence in the radio activity of the BCGs in our sample. 

\begin{figure}
\plotone{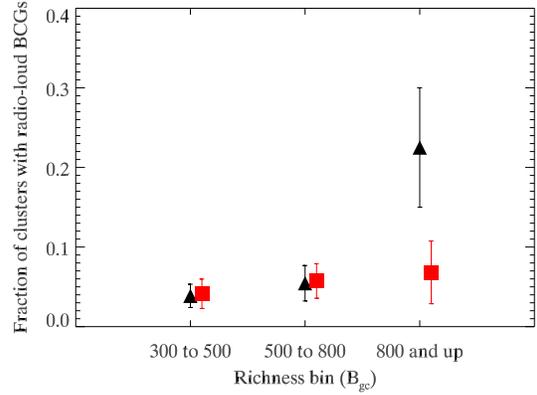} 
\caption{The fraction of galaxy clusters with radio-loud BCGs, as identified visually.  The black 
triangles show the fraction for the low redshift sample, and the red squares show the fraction for the 
high redshift sample.  \label{bcgplot}}
\end{figure}

Figure \ref{bcgplot} shows the fraction of galaxy clusters, divided into bins according to richness 
and redshift, that contain radio-loud BCGs.  Poisson error bars are plotted.  We find that the fraction of 
radio-loud BCGs does not depend strongly on richness for most of the sample.  However, the prevalence of radio-loud 
BCGs for clusters in the low-redshift, richest bin is higher than the number of radio sources per 
cluster in the other bins, possibly indicating evolution in the radio activity of the BCGs
 in the richest clusters.  We calculate the uncertainty on our measurement of the  difference 
between the fraction of radio-loud BCGs in the high redshift bin compared to the low redshift bin by adding in 
quadrature the uncertainties for each bin.  We find that the radio-loud BCG fraction in the richest low-redshift clusters
is larger than that in the richest high-redshift clusters with a significance at the 1.9$\sigma$ level.
When we compare the radio-loud BCG fraction for the richest low redshift clusters to the radio-loud BCG
fraction of all other clusters, the excess is significant at the 2.2$\sigma$ level. 
The numbers of clusters in the
 richest bins are also comparatively small (9 out of 40 clusters have radio-loud BCGs at low redshift and 3 out of 
44 clusters have radio-loud BCGs at high redshift.)  
However, nearby massive clusters are the most likely to have short 
cooling times with respect to the Hubble time, potentially allowing 
cool gas from the ICM to accrete onto the BCG, and possibly onto the central AGN.  
This picture is 
supported by recent observational studies that find a lack of strong cooling core clusters at higher redshifts 
\citep[such as $z > 0.5$ and $z > 0.7$, ][]{vikhlinincc,santos}.
If this scenario holds, then we would expect an excess of radio-loud BCGs in the richest nearby clusters
in comparison with less massive or higher redshift clusters, which is what we observe.  
 Further studies on larger samples will be necessary to confirm or disprove this excess.

\subsubsection{Non-BCG Radio Galaxies}

\begin{figure}
\plotone{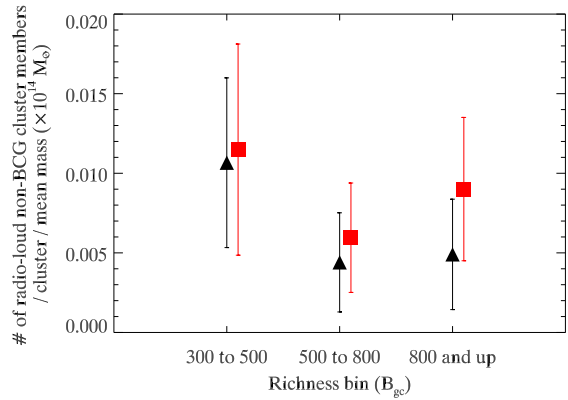}
\caption{The number of radio-loud non-BCG cluster members 
 as identified visually, per cluster, normalized by the average mass (derived from richness) for the sample of clusters
in each bin.  The black triangles show the fraction for the low redshift sample, and the 
red squares show the fraction for the high redshift sample.  \label{nobcgplot}}
\end{figure}

We identified every radio source whose optical counterpart is an early-type galaxy with a color consistent with 
the red sequence of its galaxy cluster but that is not identified as the cluster's BCG.
Figure \ref{nobcgplot} shows the number of such galaxies per cluster, divided into bins according to richness
 and redshift, normalized by the mean cluster mass (as derived from richness) for each bin.  
The number of non-BCG galaxies in a cluster is proportional to its richness, so this is similar to the 
radio loud fraction for these clusters.  Some 
of these clusters also contain radio-loud BCGs, but as long as there is also 
an excess of radio-loud galaxies in addition to (or instead of) the BCG, the radio sources are included 
in this plot.  These cluster member candidates 
were identified based on the color and magnitude of their identified optical counterparts placing them 
on the red sequence as well as their optical morphology.  Unlike in the statistical analysis, if two components
of an extended source are identified as belonging to the same optical counterpart, the system is counted
as one radio source in this plot.
For each richness bin, the number of non-BCG radio sources per cluster (normalized by mass) is somewhat higher
for the high redshift sample than for the low redshift sample.  

Thus, strong evolution in the radio-activity in one of those populations (BCG versus cluster galaxy) is not masking 
strong evolution in the 
other population.  Also, because we are matching radio sources with optical sources by eye, we do not count
radio sources that are resolved into multiple components in FIRST as separate from each other.  Morphological
evolution such that the resolution into components is different for the high and low redshift samples
could mock evolution in the radio-loud population in our statistical sample.  Thus, this visual identification
of radio sources with BCGs and cluster members upholds the overall view that there is little
 evolution in the population of radio sources in clusters from redshift 0.95 to redshift 0.35, although the
2-$\sigma$ indication of an excess of radio-loud BCGs in the richest low-redshift clusters is interesting
evidence for evolution that is worth exploring with a larger data set. 

\subsubsection{Multiple Radio Sources}\label{multipleradiosources}

The FIRST catalog splits multiple components of extended radio sources into different sources, so to explore the frequency of 
multiple radio active galaxies within clusters, it is instructive to view them by eye.  For example,
the radio source in Figure \ref{prettypic} was identified as two sources in the FIRST catalog.
We have identified twelve extended radio sources in our galaxy cluster sample that
 are split into multiple components by FIRST.  These multiple components contaminate our statistical
analysis, in which we consider radio sources as individual and uncorrelated, although the contamination
is somewhat alleviated because we treat multiple components as separate radio sources both in the
determination of the background level and in the determination of the signal.  

Twelve galaxy clusters have multiple radio sources projected within 0.5 Mpc from their centers that we have 
identified as distinct from each other, two of which have three apparently independent radio sources.  
Fourteen percent ($\pm 2\%$ Poisson counting uncertainties) of the galaxy clusters in our sample have radio 
sources within a projected 0.5 Mpc of their centers (including sources we expect from the background level.)  
Of these galaxy clusters, 14$\pm 4\%$ have multiple radio sources identified as independent sources, 
and 17$\pm 12\%$ of the sources with
multiple radio sources identified as independent have three independent radio sources.  Because all of
these percentages agree with one another to within their uncertainties, we note that galaxy clusters with one radio source are
not much more or less likely to have a second radio source, and galaxy clusters with two radio sources
are not much more or less likely to have three.  In other words, the radio loudness of galaxy cluster
members seems to be independent of the radio loudness of other galaxy cluster members, although we note
that with such a small number of clusters in the sample with multiple independent sources, the constraints 
are weak.  

One of the strengths of our statistical analysis is that radio sources which are actually lobes of a 
central radio galaxy are included in the sample.  Because we include all sources within 0.5 Mpc of 
the center of the cluster, we are likely including sources of bright radio emission which are 
associated with the BCG even if they are not coincident with the BCG because they are lobes.  Visual 
inspection of the radio sources and their relationship to the optical galaxies gives us an idea of 
how important this population is.  Of the 37 radio-loud BCGs we identify in our sample, 9 of them 
have extended radio morphology.  Of these, 6 are likely FRI's and 3 FRII's \citep{fr} based on their morphologies.
For a recent study comparing the environments of extended and point-like radio sources, see \citep{wing}.
We also calculated the luminosities of these extended sources from their NRAO VLA Sky Survey 
\citep[NVSS; ][]{nvss} fluxes.  The NVSS was conducted at 1.4 GHz with the VLA in its compact D and DnC configurations,
so is better suited to measuring the fluxes of extended sources than FIRST, which may resolve out some
flux.  \citet{ledlow} found that for Abell clusters, the dividing line in the
radio luminosity--R-band magnitude plane between FRI and FRII radio galaxies, which is very 
sharp in the general radio galaxy population \citep[although this sharpness has been recently disputed, see ][]{linmorphologies}, 
is not as distinct in nearby Abell clusters.
In our sample, all 6 FRI's lie below the line as expected, and 2 of the 3 FRII's lie above
the line as expected, but one FRII is less radio-luminous than expected.  However, \citet{ledlow} 
describe the FRII's that lie within the FRI part of the radio luminosity--magnitude plane 
as being unusually small, with the lobes confined to the optical extent of the galaxies that host them.
The only low-luminosity FRII identified in our sample looks fairly typical, with lobes located
$\sim$ 200 kpc from the host galaxy.
Larger samples are needed to better investigate the prevalence of FRII's, both luminous and underluminous
with respect to the FRI/II luminosity division, in clusters at high redshift.

There were no occasions in this analysis where one radio source was within 0.5 Mpc of two cluster 
centers.

\subsubsection{Identification of Foreground and Background Sources}

We inspected all of the galaxy clusters that contain radio sources within 0.5 Mpc of their centers.
  Some of these radio sources are chance projections.  Given the background estimates from our statistical
analysis, we expect $36.2 \pm 0.3$ radio sources within 0.5 Mpc of clusters' centers by random superpositions.
Radio sources are matched by eye with optical counterparts.  If the optical counterparts are 
elliptical galaxies with colors consistent with the red sequence at the redshift of the potentially
associated cluster, the radio source is identified as a cluster member.  Thirty-two galaxy clusters contain radio 
sources within the projected inner 0.5 Mpc that are not identified with cluster members.  Of these, seven have multiple sources
(2 of the galaxy clusters' multiple sources are identified as independent, the others are identified as multiple components.)
The consistency of these numbers provides a check that we are being 
neither too generous nor too miserly in our identification of radio loud cluster members, because the number 
of galaxies we leave out as either foreground, background or indeterminate is consistent with what we expect
from the background level.  This consistency also indicates that most of the bright radio sources
associated with galaxy cluster members are hosted by early type galaxies that lie on the red sequence.

\section{Discussion}

\subsection{Comparison with Previous Work} \label{xray}

\subsubsection{Radio Galaxy Fraction}

We compare the number of radio sources per cluster found in our sample with
the number of radio sources per cluster found in \cite{branchesi}, who studied radio AGN in X-ray selected galaxy clusters
with $0.3 \le z \le 0.8$.
The FIRST survey has a higher sensitivity limit than their pointed observations,
 although they also used the VLA operating in the B configuration, so their resolution is the same.  
They also assume a spectral index of 0.8, and they include radio sources within 0.4 Mpc of the clusters' centers. 
Their cluster sample is X-ray selected from 87.4 deg$^2$ of sky and contains 18
clusters with $0.3 < z < 0.8$, although only one cluster has $z > 0.7$.  In order to compare, we
restrict our sample to the most massive (B$_{gc,R} > 800$) clusters with $0.35 < z < 0.69$.  There
are 50 clusters in this richness and mass range, so we are probing to lower cluster mass than 
\cite{branchesi}, who have 18 clusters in about twice the area of sky but over a similar redshift range,
although because their sample of clusters is X-ray selected, their sensitivity as a function of redshift
is different than for the optically selected sample.  
When we calculate the number of excess radio sources
with luminosities greater than 1 $\times 10^{25}$ W Hz$^{-1}$ within 0.4 Mpc of these cluster centers,
we find the number per cluster to be $0.18 \pm 0.06$.
According to their Figure 8, they find approximately $0.1 \pm 0.08$ radio sources with such bright
luminosities per cluster.  Thus, the studies are in general agreement with each other on the
number of radio sources in clusters in this redshift range.

In order to compare to an optically-selected galaxy cluster sample, we also compare our data to the 
results of \cite{croft} who correlate brightest cluster galaxies from the 
maxBCG catalog of galaxy clusters in the Sloan Digital Sky Survey with FIRST radio sources.  
They use an algorithm to identify radio sources with double lobes that are likely associated with the BCG, so
their radio source sample includes both point sources and extended sources.
They have a much larger sample (over 13,000 clusters) but are restricted to low redshift ($0.1 < z < 0.3$).
  A direct comparison is not straightforward because their 
luminosity cutoff is low (1 mJy at $z = 0.1$ corresponds to about $2.5 \times 10^{22}$ W Hz$^{-1}$) and the 
mean cluster N$_{200}$ richness parameter for this sample is 17.7, corresponding to 
approximately $1 \times 10^{14}$ M$_{\odot}$ \citep{johnston}.  
We compare the RCS1 clusters that fall between redshift 0.35 and 0.45 with their sample.  
Using a luminosity cutoff of $6.5 \times 10^{23}$ W Hz$^{-1}$ (which corresponds to a 
1.0 mJy radio source at a redshift of 0.45)
we find the number of radio sources per RCS1 cluster in excess of the background of radio sources to 
be 0.29 $\pm$ 0.06.  We then scale this number to account for the differences in average mass, in radio luminosities probed, 
and in the fact that we are counting all radio-loud galaxies in clusters, not just BCGs.  
The average B$_{gc,R}$ of this RCS1 subsample is 627, corresponding to $4.3 \times 10^{14}$ M$_{\odot}$. 
Extrapolating from the field 1.4 GHz luminosity function for radio AGN in \citet{fieldrlf} (\citet{ledlow} find no difference
between the cluster and field radio luminosity functions) we expect the radio sources in 
\citet{croft} to be about 3.4 times as numerous as the radio sources to which we are matching RCS1 
clusters in this comparison.  Finally,
from our identification of radio-loud BCGs as discussed in section \ref{bcgs}, we find that 75\% of clusters
with radio sources above the level expected from the background have radio-loud BCGs.  Thus the scaled radio-loud BCG fraction
for these RCS1 clusters is $17.4\%$.  The Poisson uncertainty on this fraction is about 4\%, but the uncertainties in extrapolating across cluster
masses and the radio luminosity function are probably more significant. \cite{croft} find that 19.7\% ($\pm0.4\%$ Poisson 
uncertainty) of their BCGs are radio-loud.  
Thus without accounting for 
any possible redshift evolution, and with the above extrapolations for differences in cluster and radio
source properties, the two studies approximately agree.

\subsubsection{Redshift Evolution of Radio Galaxy Fraction}

Most of the previous work on the evolution of radio sources in clusters has been based on in-depth studies of relatively
small samples of intermediate to high redshift X-ray selected galaxy clusters in comparison to in-depth
studies of nearby Abell clusters.    

\cite{stocke} conducted pointed 1.4 GHz observations of 19 $0.3 < z < 0.8$ X-ray selected galaxy clusters.
They compared the radio activity of galaxies in these clusters to the radio activity observed in 
nearby clusters by \cite{ledlow} and found no difference between the two samples.  Extending this work 
to progressively higher redshift, \cite{perlman} conducted pointed 1.4 GHz observations of 17 $0.5 < z < 1$
X-ray selected galaxy clusters.  They also found no evidence for evolution with redshift in the cluster
radio source population.

\cite{branchesi} calculated the radio luminosity function of 18 X-ray selected galaxy clusters with 
$0.3 \le z \le 0.8$.  They compared their results to the low-redshift studies of \cite{fanti} and 
\cite{ledlow} and concluded that their radio luminosity function is significantly higher than the RLF for 
nearby galaxy clusters.  However, as they discussed in their conclusions, their X-ray selected clusters
have, on average, higher mass than the local Abell clusters to which they are comparing, and when they
take into account the presumed difference in richness between the two samples, as \cite{stocke} do, 
they no longer discern any difference in the amplitude of the radio luminosity function between the 
high and low redshift samples.

Thus, when the studies of radio AGN in high-redshift, X-ray selected clusters properly account for the average 
cluster richness of their high redshift samples,
they also see no evidence for evolution of the radio-loud cluster galaxy population with redshift, just as
we see no evidence for evolution using our uniformly selected sample (see section \ref{stats}).  Of course, 
because these other studies use nearby Abell 
clusters as their low redshift clusters, they are probing a more local cluster population than we 
do, and if the radio AGN in clusters evolve strongly between a redshift of 0 and a redshift of 0.3, 
then these other studies would be sensitive to evolution in a redshift range for which we have no data.
However, the previous studies have few clusters with redshifts $> 0.6$, so we are more sensitive 
to evolution at the higher redshift range.  

\cite{bestevolution} conducted deep radio observations of a rich $z \sim 0.83$ galaxy cluster, MS1054--03, 
finding 8 spectroscopically confirmed cluster members with radio emission brighter than 32$\mu$Jy, a 
large excess compared
when compared with low redshift clusters.  Such deep radio observations probe 
a different part of the luminosity function, which at low luminosities is dominated by star forming
galaxies rather than AGN.  
Instead, we focus on the brighter,
rarer AGN over a larger statistical sample.

\subsection{Implications}

\subsubsection{The Effect on SZ Cluster Surveys}

\cite{linmohr} generate predictions of point source contamination for SZ surveys.  The evolution of 
the radio source population is one of their unknowns, so they investigate two models: no evolution, 
and a power-law model in which the number of radio AGN in a cluster of mass M scales as ($1+z$)$^{\gamma}$
where $\gamma = 2.5$.  For this latter scenario, and assuming their propogation of a spectral
index distribution to determine the predictions of fluxes at higher (30, 90, 150 GHz) frequencies
from their counts at 1.4 GHz, they determine the fraction of clusters in an SZ survey that would be lost
due to contamination from radio sources as a function of cluster mass and redshift.  Because the SZ flux
is roughly constant with redshift, and because of the increasing luminosity distance with redshift (and 
with consideration of the $k$-correction), higher luminosity radio sources are required to contaminate
the SZ effect from clusters at higher redshifts.  However, with their power-law redshift evolution, they
find that for low-mass clusters, the evolution of the radio source population wins out over these other
effects so that the fraction of lost clusters is higher at high redshift than at low redshift.  Conversely, for
high mass clusters with greater SZ signals, and because of the steepness of the radio luminosity function,
the fraction of lost clusters is higher at low redshift than at high redshift.  
With no evolution, the lost cluster fraction would be decreased for each redshift bin by a factor 
of ($1+z$)$^\gamma$, which for the redshifts they consider (0.1, 0.6, 1.1) when looking at a range in mass
(their Figure 15) corresponds to factors of $\sim$ 1.3, 3.2 and 6.4.  

We do not see evolution in the cluster radio source population, and if anything, there is some small evidence
(1.5$\sigma$) that lower redshift clusters have more radio sources per unit mass than high redshift clusters.
In order to quantify how significantly our data constrain this type of redshift evolution, we have performed 
a chi-square test.  
We calculated the number of radio sources expected for ($1+z$)$^{\gamma}$
evolutionary models, using the cluster redshifts and normalizing to the total number of excess radio sources in clusters
that we observe in the real data.  We compare the models with the data for the number of radio sources per cluster for 
the low and high redshift bins (without accounting for the richness dependence other than by applying a uniform richness cut).  
The preferred value of $\gamma$ is $-1.9 \pm 2.8$,
where the $3\sigma$ uncertainties given are estimated from fitting a Gaussian to the P-values as a function of $\gamma$.
This test also indicates that there is a $0.7\%$ probability that the difference between our data and
a $\gamma=2.5$ evolutionary model is purely random, implying that such evolution is unlikely.

Qualitatively, the main change, based on our study, in the forecasts from \citet{linmohr} for SZ survey contamination
given no evolution in the cluster-associated radio source population is that the low mass, high redshift 
clusters are not as badly contaminated as low mass, low redshift
clusters; so, fewer high redshift clusters would be lost from SZ surveys.  In their more recent study, which also includes
observations of the spectral indices of radio sources in clusters, \citet{bruce} adapted a more mild evolution ($\gamma = 1$), motivated
by this work.  They found that this makes the SZ contamination by AGN much smaller than had been previously estimated.

\subsubsection{Comparison with Cluster X-ray and Infrared AGN Frequency}

Studies that focus on X-ray selected AGN in galaxy clusters observe a strong evolution over a redshift
range similar to what we probe.  \cite{martini} observe a dramatic increase (by a factor of eight) in the population
of X-ray selected AGN in galaxy clusters over a redshift range of 0.05 to 1.3.  
\citet{galametz} conclude that for X-ray-selected, optically-selected and radio-selected AGN, the enhancement
in AGN activity in the centers of clusters is greater at $0.5 < z \le 1$ than at $z \le 0.5$. Although the field-corrected
radio AGN density remains consistent between the redshift bins to within their quoted uncertainties, they do 
see some low significance evidence for evolution in the radio source population by about a factor of two.  The clusters
in their study are selected from optical and infrared imaging data.  The evolution 
in the X-ray AGN population drives their conclusions.

Our data rule out strong evolution of the radio-loud AGN population
between a redshift of 0.35 and 0.95 in our sample.  The two AGN populations (radio and X-ray) may be evolving differently from each
other in the cluster environments.  
Evidence of evolution in the X-ray AGN population does not imply evolution in the cluster radio AGN population.
\cite{quyen} identified X-ray AGN and radio AGN in 11 galaxy clusters at $0.2 < z < 0.4$ selected from X-ray data as 
clusters that are similar to what the Coma cluster would have been like at that redshift.  They found that X-ray 
point sources and radio sources rarely overlap in their sample. 
The two populations are separate, and so predictions that rely on 
the radio source population evolution (such as projections for the contamination of SZ surveys) should
not use the observed cluster X-ray AGN evolution.  

Studies of X-ray AGN in clusters typically define a cluster galaxy AGN fraction as their observable quantity.  They 
typically exclude the BCGs from this AGN fraction to avoid possible contamination from cluster cool cores \citep[e.g., ][]{martini06, arnold}.
Of the different results for radio sources in clusters that we have presented, the quantity that we measure that compares most directly
to this AGN fraction is the number of radio-loud non-BCG cluster members per unit mass (as shown in Figure \ref{nobcgplot}).
While the slight increase at high redshifts 
is not significant (the significance of the difference between the mean number of non-BCG radio sources per cluster per 
unit mass for the high redshift bins and for the low redshift bins is only 1.2$\sigma$), the redshift range we are probing is also 
not as wide as that probed by \cite{martini}.   
They measure the fraction of galaxies in clusters that are X-ray AGN using two different redshift binning schemes
with two and three redshift bins, and restricting their sample to the two highest redshift bins of the three bin 
scheme brings it into better agreement with our redshift distribution (these bins have $0.3 < z < 0.6$ and $z > 0.6$).
The mean X-ray AGN fraction they measure for these bins are 0.0031 and 0.0147, so that the increase is about a factor of  
4.7.  Such a large increase is unlikely in the radio source population in our sample.  
So even if qualitatively there may be some similar
evolution (at quite low significance) in the radio population as the X-ray population, it appears to be less strong.
Thus X-ray AGN and radio-loud AGN are likely two different AGN populations, and they show different redshift evolution
in galaxy clusters.

\section{Conclusions}

We constrain the evolution of the bright central radio source population in galaxy clusters from redshift 
0.35 to 0.95 by statistically matching FIRST radio sources with 618 galaxy clusters from a uniformly, optically selected sample (RCS1).   
The number of radio sources per cluster (0.14 $\pm$ 0.02 for clusters with $0.35 < z < 0.65$ and 0.10 $\pm$ 0.02 for clusters with $0.65 < z < 0.95$)
is consistent with previous studies using different, more local samples. 
Richer clusters have more radio sources per cluster, also as expected from previous studies.  After normalizing 
by the clusters' mass (as derived from richness), we find that the number of radio sources 
per unit cluster mass is roughly constant with mass and with redshift (the significance of the difference between high and low redshift clusters
is 1.5$\sigma$).
The number of radio sources (with L$_{1.4 GHz} > 4.1 \times 10^{24}$ W Hz$^{-1}$) per $10^{14}$ M$_{\odot}$ of cluster mass is 0.031$\pm0.004$. 

We also visually inspect the radio source emission in the galaxy clusters with radio sources in our sample.
Identifying BCGs and matching radio emission with optical objects, we find some (2-$\sigma$) evidence that the number of
radio-loud BCGs is higher for high richness, low redshift clusters than for either high richness,
high redshift clusters or low richness (at high or low redshift) clusters.  If there is a connection between the radio activity 
of the BCG of a galaxy cluster and amount of cooling in the surrounding ICM and if the lowest redshift,
richest clusters are most likely to have large amounts of cooling gas in their inner regions, then
we might expect to observe more radio-loud BCGs in nearby rich clusters.  
The low and intermediate richness bins do not show any evolution in the prevalence of radio-loud BCGs.

Studies of the cluster X-ray AGN population find strong evolution over a similar redshift range as we probe, while
we do not observe such evolution in the radio AGN population.  This indicates that X-ray AGN and radio-loud
AGN likely probe different AGN populations.

The RCS collaboration is currently working on a second, larger cluster catalog based on imaging 
almost 1000 square degrees in three filters (g', r', z') with the CFHT MegaCam.  We plan on 
extending this statistical analysis of the radio source population in galaxy clusters when the new 
survey becomes available in order to further probe the effects of galaxy cluster properties on the radio source
population and to further characterize the cluster radio sources.

\begin{acknowledgements}
The RCS1 project is partly supported by grants to HKCY from the Natural Science and Engineering Research Council of Canada and the
Canada Research Chair program.  The data used in this paper are based on observations obtained at the Canada-France-Hawaii Telescope (CFHT) which is operated by the National Research Council of Canada, the Institut National des Sciences de l'Univers of the Centre National de la Recherche Scientifique of France, and the University of Hawaii.
LFB is partly supported by the Chilean Centro de Astrofisica FONDAP No. 15010003, and by CONICYT
under project No. 1085286.
\end{acknowledgements}

{\it Facilities:} \facility{VLA}, \facility{Blanco}, \facility{CFHT}

\end{document}